\documentclass[prl,twocolumn]{revtex4}
\usepackage[dvips]{epsfig}
\usepackage{amsmath,amssymb}

\parskip=\medskipamount

\def\be{\begin{equation}}
\def\ee{\end{equation}}

\begin{document}

\title{An Anisotropic Ballistic Deposition Model with Links to the Ulam Problem
and the Tracy-Widom Distribution}

\author{Satya N. Majumdar$^{1}$ and Sergei Nechaev$^{2,3}$}

\affiliation{$^1$Laboratoire de Physique Quantique (UMR C5626 du CNRS),
Universit'e Paul Sabatier, 31062 Toulouse Cedex, France \\ $^2$LPTMS,
Universit\'e Paris Sud, 91405 Orsay Cedex, France \\ $^3$Landau Institute for
Theoretical Physics, 117334 Moscow, Russia}

\date{\today}

\begin{abstract}  
We compute exactly the asymptotic distribution of scaled height in a
(1+1)--dimensional anisotropic ballistic deposition model by mapping it to the
Ulam problem of finding the longest nondecreasing subsequence in a random
sequence of integers. Using the known results for the Ulam problem, we show
that the scaled height in our model has the Tracy-Widom distribution appearing
in the theory of random matrices near the edges of the spectrum. Our result
supports the hypothesis that various growth models in $(1+1)$ dimensions that
belong to the Kardar-Parisi-Zhang universality class perhaps all share the same
universal Tracy-Widom distribution for the suitably scaled height variables. 
\medskip

\end{abstract}

\maketitle

Growth processes are ubiquitous in nature. The past few decades have seen an
extensive research on a wide variety of both discrete and contiuous growth models
\cite{Meakin,KS,HZ}. A large class of these growth models such as the Eden model
\cite{Eden}, restricted solid on solid (RSOS) models \cite{RSOS}, directed
polymers \cite{HZ}, polynuclear growth models (PNG) \cite{PNG} and ballistic
deposition models (BD) \cite{BD} are believed to belong to the same
universality class as that of the Kardar-Parisi-Zhang (KPZ) equation describing the
growth of interface fluctuations \cite{KPZ}. This universality is, however,
somewhat restricted in the sense that it refers only to the width or the second
moment of the height fluctuations characterized by two independent exponents
(the growth exponent $\beta$ and the dynamical exponent $z$) and the associated
scaling function. Moreover, even this restricted universality is established
mostly numerically. Only in very few special discrete (1+1)-D models, the
exponents $\beta=1/3$ and $z=3/2$ can be computed exactly via the Bethe ansatz
technique \cite{Bethe}. A natural and important question is whether this
universality can be extended beyond the second moment of height fluctuations.
For example, does the full distribution of the height fluctuations (suitably
scaled) is universal, i.e. is the same for different growth models belonging to
the KPZ class? Moreover, the KPZ-type equations are usually attributed to 
models with small gradients in the height profile and the question whether the
models with large gradients belong to the KPZ universality class is still open.

Recently Pr\"ahofer and Spohn \cite{PS} found an exact mapping between a
specific PNG model and the so-called longest increasing subsequence (LIS)
problem, also known as the Ulam problem. The LIS problem was first raised by
Ulam in the early 60's \cite{Ulam}, then the interest in it reappeared in the
mathematical literature in 70's since the work of Vershik and Kerov \cite{VK}.
The exact mapping of PNG to LIS and the subsequent utilization of the exact
results available for the LIS problem allowed Pr\"ahofer and Spohn to find
(besides the exact KPZ growth exponent $\beta=1/3$) the exact asymptotic height
distribution in the PNG model \cite{PS}. This distribution turned to be the
well known Tracy-Widom distribution appearing in the theory of edge states of
random matrices \cite{TW}. Around the same time, Johansson showed rigorously
\cite{Johansson} that a specific (1+1)-D directed polymer model, believed to be
in the KPZ universality class, also has the same Tracy-Widom distribution for
the scaled height (energy) variable. Gravner et. al. found the same Tracy-Widom
distribution in another class of $(1+1)$ dimensional growth models which they called
`oriented digital boiling' model\cite{GTW}.  
It would be interesting to know whether
there are other growth models such as the RSOS or the BD ones, which are
believed to be in the KPZ universality class as far as the second moment is
concerned, also share the same Tracy-Widom distribution for the scaled height. 

The purpose of this Letter is to present a BD model which can be mapped
exactly to the LIS problem and hence it shares the same Tracy-Widom
distribution as the PNG model. This exact result, in combination with the
results of \cite{PS,Johansson,GTW}, then lends support to the hypothesis that
perhaps all these different growth models, at least in (1+1) dimensions,
share the same universal Tracy-Widom distribution for scaled height. This
hypothesis, if true, puts the universality on a much stronger footing going
beyond just the second moment. Incidentally, to our knowledge, our model is the
first exact solution for the full asymptotic height distribution of BD type
systems. 

Before describing our model, it is worth summarizing the main results
for the LIS problem that we use later. Take a set of $n$ integers $\{1,2,3,\dots, n\}$.
Consider all the $n!$ possible permutations of this sequence. For any given
permutation, let us find out all possible increasing subsequences (terms of a
subsequence need not necessarily be consecutive elements) and from them find
out the longest one. For example, take $n=10$ and consider a particular
permutation $\{8, 2, 7, \underbar 1, \underbar 3, \underbar 4, 10, \underbar 6,
\underbar 9, 5\}$. From this sequence, one can form several increasing
subsequences such as $\{8,10\}$, $\{2,3,4,10\}$, $\{1,3,4,10\}$ etc. The
longest one of all such subsequences is either $\{1,3,4,6,9\}$ as shown by the
underscores or $\{2,3,4,6,9\}$. The length $l_n$ of the LIS 
(in our example $l_n=5$) is a random
variable as it varies from one permutation to another. In the Ulam problem one
considers all the $n!$ permutations to be equally likely. Given this uniform
measure over the space of permutations, what is the statistics of the random
variable $l_n$? Ulam found numerically for the average length $\langle
l_n\rangle$ the asymptotic behavior $\langle l_n\rangle\sim c \sqrt{n}$ for
large $n$. Later this result was established rigorously by Hammersley
\cite{Hammersley} and the constant $c=2$ was found by Vershik and Kerov
\cite{VK}. Recently, in a seminal paper, Baik, Deift and Johansson (BDJ)
\cite{BDJ} derived the full distribution of $l_n$ for large $n$. In particular,
they showed that $l_n \to 2\sqrt {n} + n^{1/6} \chi$ for large $n$, where the
random variable $\chi$ has an $n$-independent distribution which happens to be
the Tracy-Widom distribution for the largest eigenvalue of a random matrix
drawn from the Gaussian Unitary Ensemble \cite{TW}. They also showed
that when the sequence length $n$ itself is a random variable drawn from a
Poisson distribution with mean $\langle n\rangle =\lambda$, the length of the LIS converges for
large $\lambda$ to 
\be
l_{\lambda}\to 2\sqrt{\lambda} + {\lambda}^{1/6} \chi, 
\label{bdj1} 
\ee 
where $\chi$ has the Tracy-Widom distribution. The fixed $n$ and the fixed
$\lambda$ ensembles are like the canonical and the grand canonical ensembles in
statistical mechanics. The detailed form of the Tracy-Widom distribution is
rather complicated and not very illuminating (see \cite{PS} for a picture). The
BDJ results led to an avalanche of subsequent mathematical works \cite{AD}. 

\begin{figure}
\centerline{\epsfig{file=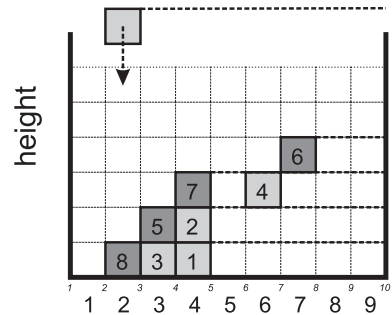,width=5cm}}
\caption{Growth of a heap with asymmetric long-range interaction. The numbers
inside cells show the times at which the blocks are added to the heap.} 
\label{fig:1}
\end{figure}

In our (1+1)-D BD model columnar growth occurs sequentially on a linear substrate
consisting of $L$ columns with free boundary conditions. The time $t$ is
discrete and is increased by $1$ with every deposition event. One starts at
$t=0$ with an empty substrate. At any stage of the growth, a column (say the
$k$-th column) is chosen at random with probability $p=\frac{1}{L}$ and a
"brick" is deposited there which increases  the height of this column by one
unit, $H_k\to H_k+1$. Once this "brick" is deposited, it screens all the sites
at the same level in all the columns to its right from future deposition, i.e.
the heights at all the columns to the right of the $k$-th column must be
strictly greater than or equal to $H_k+1$ at all subsequent times. For example,
in Fig.\ref{fig:1}, the first brick (denoted by 1) gets deposited at $t=1$ in
the 4-th column and it immediately screens all the sites to its right. Then the
second brick (denoted by 2) gets deposited at $t=2$ again in the same 4-th
column whose height now becomes 2 and thus the heights of all the columns to
the right of the 4-th column must be $\ge 2$ at all subsequent times and so on.
Formally such growth is implemented by the update rule,
\be 
H_k(t+1)={\rm max}\{H_k(t), H_{k-1}(t), \dots, H_1(t)\}+1. 
\label{update1} 
\ee 
The model is anisotropic and evidently even the average height profile $\langle
H_k(t) \rangle$ depends nontrivially on both the column number $k$ and time
$t$. Our goal is to compute the asymptotic height distribution $P_k(H,t)$ for
large $t$. 

It is easy to find the height distribution $P_1(H, t)$ of the first column,
since the height there does not depend on any other column. At any stage, the
height in the first column either increases by one unit with probability
$p=\frac{1}{L}$ (if this column is selected for deposit) or stays the same with
probability $1-p$. Thus $P_1(H,t)$ is simply the binomial distribution,
$P_1(H,t)={t\choose H}p^h(1-p)^{t-H}$ with $H\leq t$. The average height of the
first column thus increases as $\langle H_1(t)\rangle=pt$ for all $t$ and its
variance is given by $\sigma_1^2(t)= tp(1-p)$. While the first column is thus
trivial, the dynamics of heights in other columns is nontrivial due to the
right-handed infinite range interactions between the columns. For
convenience, we subsequently measure the height of any other column with respect to the
first one. Namely, by height $h_k(t)$ we mean the height difference between the
$(k+1)$-th column and the first one, $h_k(t)=H_{k+1}(t)-H_1(t)$, so that
$h_0(t)=0$ for all $t$. 

\begin{figure} 
\centerline{\epsfig{file=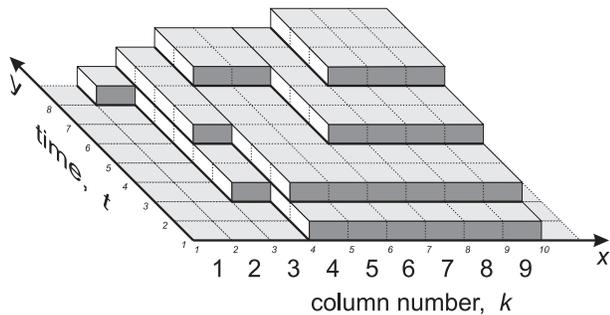,width=8cm}}
\caption{$(2+1)$ dimensional "terraces" corresponding to the growth of a heap
in Fig.\ref{fig:1}} 
\label{fig:2}
\end{figure}

To make progress for columns $k>0$, we first consider a
(2+1)-D construction of the heap as shown in Fig.\ref{fig:2}, by adding an extra
dimension indicating the time $t$. In Fig.\ref{fig:2}, the $x$ axis denotes the
column number, the $y$ axis stands for the time $t$ and the $z$ axis is the
height $h$. In this figure, every time a new block is added, it "wets" all the
sites at the same level to its "east" (along the $x$ axis) and to its "north"
(along the time axis). Here "wetting" means "screening" from 
further deposition at those sites at the same level. This $(2+1)$-D system of
"terraces" is in one-to-one correspondence with the $(1+1)$-D heap in
Fig.\ref{fig:1}. This construction is reminiscent of the 3D anisotropic
directed percolation (ADP) problem studied by Rajesh and Dhar \cite{RD}. Note however,
that unlike the ADP problem, in our case each row labelled by $t$ can contain
only one deposition event\cite{ADP}. 

\begin{figure} 
\centerline{\epsfig{file=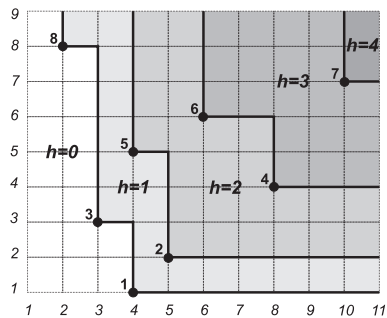,width=5cm}} 
\caption{The directed lines are the level lines separating adjacent terraces
with height diffrence $1$ in Fig. 2, projected onto the $(x,y)$ plane and
shifted by $(x,y)\to (x+h(x,y),y)$ to avoid partial overlap. The black dots
denote the deposition events. The numbers next to the dots denote the times of
those deposition events.} 
\label{fig:3}
\end{figure}

The next step is to consider the projection onto the 2D $(x,y)$-plane of the
level lines separating  the adjacent terraces whose heights differ by $1$. In
this projection, some of the level lines may overlap partially on the plane. 
To avoid the overlap for better visual purposes, we make a shift
$(x,y)\to (x+h(x,y),y)$ and represent these shifted directed lines on the 2D
plane in Fig.\ref{fig:3}. 

The black dots in Fig.\ref{fig:3} denote the points
where the deposition events took place and the integer next to a dot denotes
the time of this event. Note that each row in Fig.\ref{fig:3} contains a single
black dot, i.e. only one deposition per unit of time can occur. In
Fig.\ref{fig:3}, there are 8 such events whose deposition times form the
sequence $\{1,2,3,4,5,6,7,8\}$ of length $n=8$. Now let us read the deposition times of the
dots sequentially, but now column by column and vertically from top to bottom
in each column, starting from the leftmost one. Then this sequence reads
$\{8,3,5,1,2,6,4,7\}$ which is just a permutation of the original sequence
$\{1,2,3,4,5,6,7,8\}$. In the permuted sequence $\{8,3,5,1,2,6,4,7\}$ there are
$3$ LIS's: $\{3,5,6,7\}$, $\{1,2,6,7\}$ and $\{1,2,4,7\}$, all of the same
length $l_n=4$. There is a greedy algorithm called the ``patience sorting" game
devised by Aldous and Diaconis to determine this length of the LIS\cite{AD}. This game
goes as follows: start forming piles with the numbers in the permuted sequence
starting with the first element which is $8$ in our example. So, the number 8
forms the base of the first pile. The next element, if less than 8, goes on
top of 8. If not, it forms the base of a new pile. One follows a greedy
algorithm: for any new element of the sequence, check all the top numbers on
the existing piles starting from the first pile and if the new number is less
than the top number of an already existing pile, it goes on top of that pile.
If the new number is larger than all the top numbers of the existing piles,
this new number forms the base of a new pile. Thus in our example, we form $4$
distinct piles: $[\{8,3,1\}, \{5,2\}, \{6,4\}, \{7\}]$. The number of piles ($4$)
is same as the length $l_n=4$ of the LIS of this permuted sequence. In fact,
Aldous and Diaconis proved \cite{AD} that the length of the LIS $l_n$ is exactly
equal to the number of piles in the corresponding patience sorting game.

Let us note one immediate fact from Fig.\ref{fig:3}. The numbers
belonging to the different level lines in Fig.\ref{fig:3} are in one-to-one
correspondence with the piles $[\{8,3,1\}, \{5,2\}, \{6,4\},\{7\}]$ in
Aldous--Diaconis patience sorting game. Hence, each pile can be identified with
an unique level line. Now, the height $h(x,t)$ at any given point $(x,t)$ in
Fig.\ref{fig:3} is equal to the number of level lines inside the rectangle
bounded by the corners: $[0,0], [x,0], [0,t], [x,t]$. Thus, we have
the correspondonce: height $\equiv$ number of level lines $\equiv$ number of piles $\equiv$ 
length $l_n$ of the LIS. However, to compute $l_n$, we need to know $n$ which
is the number of black dots inside this rectangle. 

Once the problem is reduced to finding the number of black dots or deposition events, we
no longer need the Fig.\ref{fig:3} (as it may confuse due to the visual shift
$(x,y)\to (x+h(x,y),y)$) and can go back to Fig.\ref{fig:2}, where the
north-to-east corners play the same role as the black dots in Fig.\ref{fig:2}.
In Fig.\ref{fig:2}, to determine the height $h_k(t)$ of the $k$-th column at
time $t$, we need to know the number of deposition events inside the $2$D plane
rectangle $R_{k,t}$ bounded by the four corners $[0,0], [k,0], [0,t], [k,t]$. 
Let us begin with the last column $k=L$. For $k=L$ the number of deposition
events $n$ in the rectangle $R_{L,t}$ is equal to the time $t$ because there is
only one deposition event per time. In our example $n=t=8$. For a general $k<L$
the number of deposition events $n$ inside the rectangle $R_{k,t}$ is a random
variable, since some of the rows inside the rectangle may not contain a
north-to-east corner or a deposition event. The probability distribution
$P_{k,t}(n)$ (for a given $[k,t]$) of this random variable can, however, be
easily found as follows. At each step of deposition, a column is chosen at
random from any of the $L$ columns. Thus, the probability that a north-to-east
corner will fall on the segment of line $[0,k]$ (where $k\leq L$) is equal to
$k/L$. The deposition events are completely independent of each other,
indicating the absence of correlations between different rows labelled by $t$ in
Fig.\ref{fig:2}. So, we are asking the question: given $t$ rows, what is the
probability that $n$ of them will contain a north-to-east corner? This is
simply given by the binomial distribution
\be
P_{k,t}(n) = {t\choose n } \left({\frac {k}{L}} \right)^n 
\left(1-{\frac {k}{L}}\right)^{t-n}, 
\label{binom1} 
\ee 
where $n\leq t$. Now we are reduced to the following problem: given a sequence
of integers of length $n$ (where $n$ itself is random and is taken from the
distribution in Eq.(\ref{binom1})), what is the length of the LIS? Recall that
this length is precisely the height $h_k(t)$ of the $k$-th column at time $t$
in our model. In the thermodynamic limit $L\to \infty$ for $t\gg 1$ and any
fixed $k$ such that the quotient $\lambda=\frac{tk}{L}$ remains fixed but is
arbitrary, the distribution in Eq.(\ref{binom1}) becomes a Poisson distribution
$P(n)\to e^{-\lambda} \frac {\lambda^n}{n!}$, with the mean
$\lambda=\frac{tk}{L}$. We can then directly use the BDJ result in
Eq.(\ref{bdj1}) to predict our main result for the height in the BD model, 
\be 
h_k(t) \to 2\sqrt{\frac{tk}{L}} + \left(\frac{tk}{L}\right)^{1/6} \chi, 
\label{result1} 
\ee 
for large $\lambda=tk/L$, where the random variable $\chi$ has the
Tracy-Widom distribution. Using the known exact value $\langle \chi\rangle
=-1.7711...$ from the Tracy-Widom distribution\cite{TW}, we find exactly the
asymptotic average height profile in the BD model, 
\be 
\langle h_k(t)\rangle \to 2\sqrt{\frac{tk}{L}}-
1.7721...\left(\frac{tk}{L}\right)^{1/6}. 
\label{avgh} 
\ee 
The leading square root dependence of the profile on the column number $k$ has
been seen numerically \cite{details}. The Eq.(\ref{avgh}) also predicts an
exact sub-leading term with $k^{1/6}$ dependence. Similarly, for the variance,
$\sigma_k^2(t)=\langle [h_k(t)-\langle h_k(t)\rangle]^2 \rangle$, we find
asymptotically: $\sigma_k^2(t)\to c_0\left(\frac{tk}{L}\right)^{1/3}$, where
$c_0=\langle [\chi-\langle \chi \rangle]^2\rangle=0.8132...$ \cite{TW}.
Eliminating the $t$ dependence for large $t$ between the average and the
variance, we get, $\sigma_k^2(t)\approx a {\langle h_k(t)\rangle}^{2\beta}$
where the constant $a=c_0/2^{2/3}=0.51228\dots$ and $\beta=1/3$, thus
recovering the KPZ scaling exponent.
In addition to the BD model with infinite range right-handed
interaction reported here, 
we have also analyzed the model (analytically within a mean field theory and numerically)  
when the right-handed interaction is short ranged\cite{details}.
Surprisingly, we found that 
the asymptotic average height profile is independent of the range of interaction\cite{details}.

In summary, we have shown that the asymptotic scaled height in an anisotropic
(1+1)D BD model has the Tracy-Widom distribution. Our exact result, in
combination with those of Refs.\cite{PS,Johansson,GTW} where the 
same
distribution was found in rather different growth models, 
suggests that the universality in all these growth processes is perhaps much
wider than it was thought before, extending to the full asymptotic height
distribution and is not restricted only to the second moment of the height.

We thank the hospitality of the Institut Henri Poincar\'e in Paris where this
work was initiated during the trimester "Geometry and Statistics of Random
Growth" in 2003.


\begin{thebibliography}{99}
\bibitem{details} The details will be published elsewhere.
\bibitem{ADP} The ADP problem can also be mapped asymptotically to the LIS problem\cite{details}.
The ADP problem, on the other hand,  can be solved\cite{RD} by mapping it to a $5$-vertex 
model\cite{baxter}, which provides an alternative physical derivation of the BDJ results 
via the mapping to a $5$-vertex model.
\end{thebibliography}
\end{document}